\documentclass[onecolumn,amsmath,amssymb]{qm2008_abs}
\usepackage{graphicx}
\usepackage{dcolumn}
\usepackage{bm}
\begin{document}

\title{{\Large Parton Energy Loss Without Transverse Momentum Broadening}}

\bigskip

\bigskip

\author{\large Korinna Zapp$^\text{a,b}$}
\author{\large Gunnar Ingelman$^\text{c}$}
\author{\large Johan Rathsman$^\text{c}$}
\author{\large Johanna Stachel$^\text{b}$}
\author{\large Urs Achim Wiedemann$^\text{a}$}
\affiliation{$^\text{a}$ Physics Department, Theory Division, CERN, CH-1211 Geneva 23, Switzerland\\
$^\text{b}$ Physikalisches Institut, Universit\"at Heidelberg, Philosophenweg 12, D-79120 Heidelberg, Germany \\
$^\text{c}$ High Energy Physics, Uppsala University, Box 535, S-75121 Uppsala, Sweden}
\bigskip

\bigskip

\begin{abstract}
\leftskip1.0cm
\rightskip1.0cm
The \textsc{Jewel\,1.0} Monte Carlo simulates jet evolution in a medium with a microscopic description of splitting and scattering processes. In the framework of this model we investigate the transverse momentum broadening due to medium effects in different scenarios. Depending on assumptions about hadronisation, we observe either a small increase or even a slight decrease of the mean transverse momentum, but no sizeable broadening. This appears to be a natural consequence of a model
formulation which conserves energy and momentum microscopically at each splitting and at each
scattering. 
\end{abstract}

\maketitle

\section{Introduction}

It is commonly believed that interactions of a jet with the matter produced in ultra-relativistic 
nuclear collisions result in sizeable transverse momentum broadening. For instance, 
parametric estimates from models of radiative energy loss indicate, that the average 
medium-induced energy loss of the leading parton grows proportional to the square of 
the in-medium path length $\langle \Delta E\rangle \simeq \hat{q}\, L^2/2$, whereas 
$\langle p_\perp^2\rangle \simeq \hat{q}\, L$. In these models, the large numerical value of $\hat{q}$, needed to account for the measured nuclear modification factor, implies a significant  broadening 
of the jet fragmentation pattern. Also in models of collisional energy, one expects the characteristic 
$\langle p_\perp^2\rangle \propto \, L$ of a transverse Brownian motion, whose  width is
related to parton energy loss. In contrast to these {\it parametric} expectations, the strong 
suppression of leading hadrons observed experimentally is not accompanied by a visible
broadening of jet-like two-particle correlations~\cite{phenix-wp,star-wp}. 

However, these parametric estimates are obtained neglecting certain constraints. In particular, 
pQCD-motivated models of parton energy loss typically invoke a high-energy limit, in which 
energy-momentum is conserved at each interaction only approximately, up to corrections of 
order $\mathcal{O}(1/E)$.  Also, these models do not treat leading and subleading partons equally 
during the parton evolution. Hence, the description of jet-like two-particle 
correlations is affected by theoretical uncertainties, which are not present
in the description of nuclear modification factors. Moreover, it remains to be studied how 
hadronisation affects transverse momentum broadening. It is an open question whether improving on these
points in the modelling of parton energy loss can lead to a satisfactory description of 
jet-like two-particle correlations in heavy ion collisions, or whether the absence of visible
$p_\perp$-broadening in these correlations signals a deep problem with the microscopic
picture of parton energy loss on which the current models are based. 

Here, we address this question with the Monte Carlo generator \textsc{Jewel}\cite{jewel}, simulating the medium-modification of jets with exact energy-momentum conservation at each interaction vertex. 
It treats leading and subleading partons on an equal footing, and it can be interfaced with various
hadronisation mechanisms.

\section{Jet Evolution With Energy Loss (JEWEL)}

In the absence of medium effects \textsc{Jewel} simulates a standard jet evolution in vacuum. The initial parton reduces its virtuality through successive radiation and splitting processes giving rise to a parton cascade. This perturbative evolution is terminated at the infra-red cut-off $Q_0$. At this scale the parton cascade can be interfaced with a suitable hadronisation prescription, which can also be used in the presence of a medium.

We chose the parton virtuality $Q^2$ as evolution variable, because it defines the lifetime of intermediate states which can be compared directly to the mean free path in a medium. The parton cascade follows the concepts of the mass ordered cascade in the \textsc{Pythia\,6.4} event generator\cite{pythia}. The probability for a parton $a$ to evolve from an initial virtuality $Q_\text{i}$ to a lower virtuality $Q_\text{f}$ without splittings is described by a standard Sudakov form factor.

\smallskip

The scattering in a medium is characterised by the mean free path, so the parton cascade formulated in terms of momenta has to be mapped onto space-time. The space-time structure of the cascade can be constructed using the lifetimes of intermediate states. A virtual state has a lifetime of order $1/Q$ in its own rest frame, which becomes $E/Q^2$ in the rest frame of the medium. In the parton cascade, where virtual states decay into other virtual states, the lifetime of a parton with virtuality $Q_\text{f}$ that was produced in a splitting at $Q_\text{i}$ is approximately
 $\tau = \frac{E}{Q_\text{f}^2} - \frac{E}{Q_\text{i}^2}$.
The medium is regarded as a collection of scattering centres with thermal mass $m_\text{scatt} \propto T$ and density $n \propto T^3$. The scattering cross section is taken as
$	\sigma^{\rm elastic} = \int \limits_0^{\vert t_{\rm max}\vert} d\vert
t\vert\,
	\frac{\pi\, \alpha_\text{s}^2(\vert t\vert + \mu_\text{D}^2)}{s^2}
	C_\text{R}\, \frac{s^2 + (s-\vert t\vert)^2}{\left(\vert t\vert + \mu_\text{D}^2
\right)^2}$,
where $\mu_\text{D}$ is the Debye screening mass. It is assumed, that scattering does not change the virtuality of the scattered parton. The probability that a parton experiences no scattering during its lifetime is given by
	$S_{\rm no\, scatt}(\tau) = \exp\left[  - \sigma_{\rm elastic}\, n\, \tau
\right]$\, .
After a splitting the produced partons can scatter in the medium until, after a time $\tau$, the next splitting occurs.

Medium-induced radiative energy loss is included by an enhanced splitting probability $\hat{P}_{\text{a}\to \text{bc}}(z) \to (1 + f_{\rm med})\, \hat{P}_{\text{a}\to
\text{bc}}(z)$ in the parton cascade to mimic the full $2\to 3$ processes that remains to be implemented\cite{Borghini:2005em}.

\smallskip

We consider two options for hadronisation. 
First, a simple string fragmentation algorithm is interfaced with
the parton cascade at a scale of $Q_0 = 1$ GeV. This algorithm identifies the most 
energetic parton in the event and stretches a string to the nearest neighbour in momentum space until a suitable endpoint is reached, thus assuming  colour correlation between the
partons. If no endpoint can be found, an additional (anti)quark with momentum in the beam direction is added to the event (the forward and backward rapidity regions are then excluded from the following event analysis). Then, the strings are hadronised using the string fragmentation routine of \textsc{Pythia\,6.4}\cite{lund,pythia}. This prescription is much simplified as compared to state-of-the-art hadronisation routines, but is has the advantage of being directly applicable to medium-modified jets. We did not attempt to fine-tune the hadronisation prescription to data. Alternatively, local parton-hadron duality\cite{lphd} (LPHD) is implemented by continuing the perturbative evolution down to a low
hadronic scale of $Q_0 \sim 260$ MeV, and letting the resulting partons represent hadrons. 


\section{Transverse Momentum Broadening}

\begin{figure}
\centering
\input{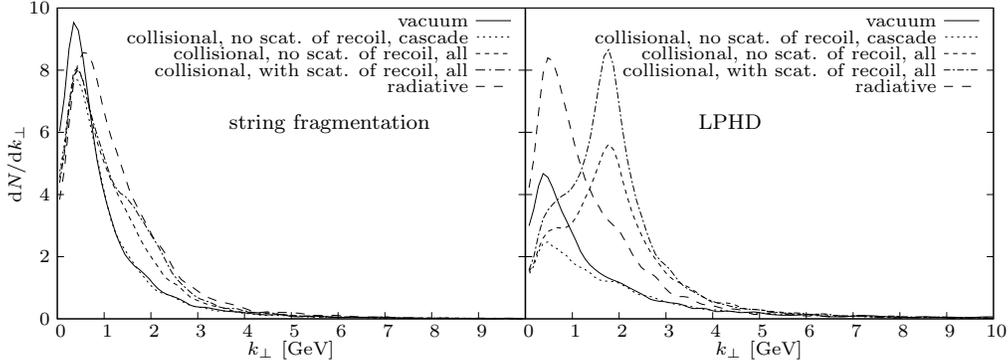}
\caption{Transverse momentum (with respect to the jet axis) distribution of hadrons with energy larger than 2\,GeV for two different hadronisation schemes and different set-ups of the elastic scattering (see text), or alternatively for radiative energy loss. The jet energy is 100\,GeV, $T=500\,\text{MeV}$, $L=5\,\text{fm}$ and $f_\text{med}=3$.}
\label{fig_kt}
\end{figure}

Fig.\,\ref{fig_kt} shows the transverse momentum distribution of all hadrons with energies larger than 2\,GeV, resulting from the fragmentation of a quark jet with $E=p_\perp =100\,\text{GeV}$. We start by discussing the fragmentation 
pattern in the absence of medium effects. The string fragmentation model 
is in rather good agreement with data: In particular, in the absence of medium effects, \textsc{JEWEL}
with string fragmentation is known to account quantitatively for the longitudinal single inclusive
distribution $\text{d}N/\text{d}\xi$, $\xi \equiv \ln E_\text{jet}/p_\text{hadron}$ within a jet~\cite{jewel}. Also, the transverse 
distribution $\text{d}N/\text{d}k_\perp$ is in reasonable semi-quantitative agreement with data 
measured in $p\bar{p}$ collisions at the Tevatron\cite{Acosta:2002gg}.
In contrast, the one-to-one mapping of partonic
to hadronic multiplicities, which underlies the LPHD model, does not provide an equally
satisfactory description of single inclusive intra-jet distributions. We noted in Ref.~\cite{jewel},
that $\text{d}N/\text{d}\xi$ from LPHD is too hard, even though the number of branchings is increased by evolving the parton cascade to $Q \sim \Lambda_{\rm QCD}$. Using this prescription, we observe (Fig.\,\ref{fig_kt}, right) that the multiplicity is reduced at low $k_\perp$; at $k_\perp \simeq 1\,\text{GeV}$ it is about 1/2 of the (realistic) results of the string fragmentation model.
For the following 
studies, the string fragmentation should be regarded as the more realistic hadronisation model. 
 It is, however, conceivable that the interactions with the medium destroy the colour correlations. 
Hadronisation may then be better represented by the LPHD option.

To discuss medium modifications, we first consider the effect of 
elastic $2\to 2$ scattering processes on the parton cascade. If the recoil partners of the medium are not included ('collisional, no scattering of recoil, cascade' in Fig.\,\ref{fig_kt}), the total multiplicity in $\text{d}N/\text{d}k_\perp$ is slightly reduced. This happens, since collisional energy loss implies that fewer hadrons pass the 2 GeV cut applied in Fig.\,\ref{fig_kt}. Remarkably, however, the broadening
of $\text{d}N/\text{d}k_\perp$ is negligible in this case. We attribute this to the fact that most elastic scatterings occur at small angle, especially for energetic
particles. When recoil particles are included in the
distribution $\text{d}N/\text{d}k_\perp$ ('collisional, no scattering of recoil, all' in Fig.\,\ref{fig_kt}), one finds an enhancement of the total multiplicity, simply
because one includes additional particles in the plot. Also, there is a small but visible
broadening effect, since most recoil partners are scattered into relatively large angle. 
The size of the contribution from recoil partners depends strongly on the low energy cut,
above which hadrons are taken into account. Most recoil partners have an energy of
less than 2 GeV~\cite{jewel} leading to only a mild broadening effect as seen in Fig.\,\ref{fig_kt}. Including multiple scatterings of the recoil partners ('collisional, with scattering of recoil, all' in Fig.\,\ref{fig_kt}) is seen again to
have a negligible effect in the case of string fragmentation, since most scatterings occur at small angle and the 'secondary' recoils have small energy. In the LPHD scenario there is an increase in mulitplicity but the shape remains unchanged.

It is also instructive to compare the two hadronisation models. As discussed already in \cite{jewel},
recoil partons are concentrated at a certain $k_\perp$ due to kinematics and the form of the 
scattering cross section, and they tend to be soft. The LPHD hadronisation model maps the 
partonic onto the hadronic distribution - thus, the characteristic shifting of the peak of $\text{d}N/\text{d}k_\perp$
is preserved by hadronisation. On the other hand, the string fragmentation model smears the peak because it produces the hadrons everywhere along the string, i.e. not only in the direction of the 
partons. As a result, the string fragmentation model leads to smaller shift in $\text{d}N/\text{d}k_\perp$. We conclude
that the characteristic peak in the (angular and $k_\perp$) recoil distribution is not stable
against assumptions about hadronisation. However, even in the LPHD scenario which preserves the peak, it does not lead to a strong increase of the mean transverse momentum.

\begin{figure}
\centering
\input{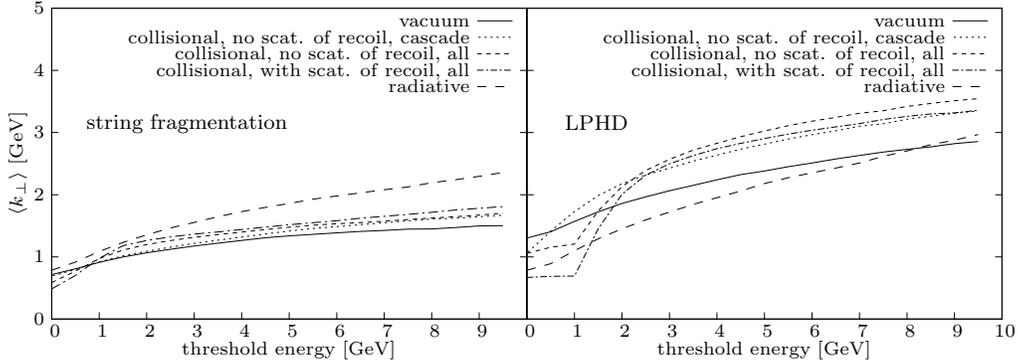}
\caption{Mean transverse momentum distribution of hadrons with energy larger than the threshold energy for the same model conditions as in Fig.\ 1.}
\label{fig_meankt}
\end{figure}

Fig.\,\ref{fig_meankt}  shows the mean transverse momentum with respect to the jet axis of particles with 
energy larger than some threshold energy. The $k_\perp$-broadening is larger in the LPHD hadronisation model, but here also the multiplicity is lower, e.g. by a factor 1.8 in the vacuum case.  
A simple scaling of the LPHD result by $\langle k_\perp \rangle_\text{h} = \langle k_\perp \rangle_\text{p}/1.8$ agrees essentially with the string model result suggesting that energy-momentum conservation is an important constraint.
Finally, the simple model of radiative energy loss explored here is seen to increase the 
jet multiplicity, and it leads to a slightly more significant jet broadening in the case of string fragmentation. This is consistent
with our findings in Ref.~\cite{jewel}.
In the LPHD scenario enhanced radiation leads to a strong increase of the multiplicity and accompanied by a reduction of the mean $k_\perp$. The string fragmentation is more efficient in converting a soft but wide distribution of partons into a hadronic distribution with increased transverse momentum, because it can merge gluons into a single hadron (provided they are in the same string, i.e. colour correlated).

In summary, our study indicates that energy-momentum conservation plays an important role in modelling transverse momentum broadening of jets. Our main conclusion is that, independent of assumptions about hadronisation and the details of collisional energy loss, there is no strong 
medium-induced transverse momentum broadening. 

\medskip

\begin{acknowledgments}
Korinna Zapp acknowledges support via a Marie Curie Early Stage Research Training Fellowship of the European Community's Sixth Framework Programme under contract number (MEST-CT-2005-020238-EUROTHEPHY). We also acknowledge support by the German BMBF and the Swedish Research Council. 
\end{acknowledgments}

\end{document}